\documentclass[reprint,prl,twocolumn,superscriptaddress,showpacs,showkeys,floatfix,preprintnumbers]{revtex4-1}

\pdfoutput=1

\usepackage{braket}
\usepackage{xargs}
\usepackage{extarrows}
\usepackage{mathtools}
\usepackage[english]{babel}
\usepackage[utf8]{inputenc}
\usepackage{amsmath,amssymb,braket,slashed,mathrsfs}
\usepackage{pifont}
\usepackage{MnSymbol}
\usepackage{graphicx}
\usepackage{color,hyperref}
\usepackage{multirow,array}
\usepackage{ulem}

\usepackage{amsmath}
\usepackage{graphicx}
\usepackage{booktabs}
\usepackage{multirow}
\usepackage{subfigure}
\usepackage{float}
\usepackage{bm}
\usepackage{diagbox}
\usepackage{multirow}
\usepackage{tabularx}

\usepackage{cleveref}
\newcommand{\ba}{\begin{eqnarray}}
\newcommand{\ea}{\end{eqnarray}}
\newcommand{\be}{\begin{equation}}
\newcommand{\ee}{\end{equation}}
\newcommand{\bi}{\begin{itemize}}
\newcommand{\ei}{\end{itemize}}

\newcommand{\nn}{\nonumber}

\newcommand{\innovation}{Collaborative Innovation Center of Quantum Matter, Beijing 100871, China}
\newcommand{\chep}{Center for High Energy Physics, Peking University, Beijing 100871, China}
\newcommand{\pkuphy}{School of Physics and State Key Laboratory of Nuclear Physics and Technology, Peking University, Beijing 100871,
China}

\newcommand{\cyi}{Computation-based Science and Technology Research Center, The Cyprus Institute, 20 Kavafi Str., Nicosia 2121, Cyprus}
\newcommand{\cyp}{Department of Physics, University of Cyprus, P.O. Box 20537, 1678 Nicosia, Cyprus}
\newcommand{\poz}{Faculty of Physics, Adam Mickiewicz University, ul.\ Uniwersytetu Pozna\'{n}skiego 2, 61-614 Pozna\'{n}, Poland}
\newcommand{\teml}{Temple University, 1925 N. 12th Street, Philadelphia, PA 19122-1801, USA}
\newcommand{\desy}{NIC, DESY, Platanenallee 6, D-15738 Zeuthen, Germany}
\newcommand{\bon}{Institut f\"{u}r Strahlen- und Kernphysik, Rheinische Friedrich-Wilhelms-Universit\"{a}t Bonn,  Nussallee 14-16, 53115 Bonn}

\begin{document}
\title{Lattice QCD Study of Transverse-Momentum Dependent Soft Function}

\author{Yuan Li}\affiliation{\pkuphy}
\author{Shi-Cheng Xia}\affiliation{\pkuphy}
\author{Constantia Alexandrou}\affiliation{\cyp}\affiliation{\cyi}
\author{Krzysztof Cichy}\affiliation{\poz}
\author{Martha Constantinou}\affiliation{\teml}
\author{Xu Feng}\affiliation{\pkuphy}\affiliation{\innovation}\affiliation{\chep}
\author{Kyriakos Hadjiyiannakou}\affiliation{\cyi}
\author{Karl Jansen}\affiliation{\desy}
\author{Chuan Liu}\affiliation{\pkuphy}\affiliation{\innovation}\affiliation{\chep}
\author{Aurora Scapellato}\affiliation{\teml}
\author{Fernanda Steffens}\affiliation{\bon}
\author{Jacopo Tarello}\affiliation{\cyi}

\date{\today}

\begin{abstract}

In this work, we perform a lattice QCD study of the intrinsic, rapidity-independent soft function within the framework of large momentum effective theory. The computation is carried out using a gauge ensemble of $N_f=2+1+1$ clover-improved twisted mass fermion. After applying an appropriate renormalization procedure and the removal of significant higher-twist contamination,
we obtain the intrinsic soft function that is comparable to the one-loop perturbative result at large external momentum.
The determination of the nonperturbative soft function from first principles is crucial to sharpen our understanding of
the processes with small transverse momentum such as the Drell-Yan production and the semi-inclusive deep inelastic scattering.
Additionally, we calculate the
Collins-Soper evolution kernel using the quasi-transverse-momentum-dependent wave function as input.

\end{abstract}

\maketitle
{\em Introduction.} --
Understanding the structure of matter within the framework of quantum chromodynamics (QCD)
is one of the central goals of hadron and nuclear physics.
Although the study of partonic transverse momentum dependent (TMD) phenomena started few years after
QCD was proposed~\cite{Parisi:1979se}, our knowledge of TMD parton distribution functions (TMDPDFs) is still limited, 
both experimentally and theoretically (see, e.g., Refs.~\cite{Scimemi:2019cmh,Collins:2016hqq}) since, 
until recently, a systematic {\em ab initio} computation of TMDPDFs was out of reach.

Knowledge of TMDPDFs would open a new
window in our understanding of hadron structure, arising, for instance, from probing the
coupling of $k_\perp$ of a given quark with its spin~\cite{Mulders:1995dh}. 
However, these functions cannot be obtained from totally inclusive processes, 
as we need an observable in the final state carrying information on $k_\perp$, 
obtained e.g. by measuring the transverse momentum $Q_\perp$ of a lepton pair 
produced in a Drell-Yan process. Consequently, they are intrinsically 
harder to measure. Nevertheless, the future Electron-Ion Collider in the U.S.~\cite{AbdulKhalek:2021gbh} and that in China~\cite{Anderle:2021wcy} have
as one of their goals to make precise measurements of TMDPDFs,
aiming to reconstruct a three-dimensional picture of hadrons in momentum space.
As in the case of collinear PDFs, the extraction of TMDPDFs from the measured
Drell-Yan or semi-inclusive deep inelastic cross sections is possible, thanks 
to factorization theorems,
which isolate the nonperturbative physics into suitable definitions of 
TMDPDFs~\cite{Collins:1984kg,Collins:1988ig,Ji:2004wu,Ji:2004xq,Collins:2017oxh}.
Unfortunately, for distributions dependent on $k_\perp$ there appears an extra
divergence associated with the emission of gluons carrying small momenta,
which is not canceled by the real and virtual perturbative corrections.
These divergences are encoded into functions called soft functions. 
At large transverse momentum $Q_\perp\gg\Lambda_{\rm QCD}$, the soft 
function can be calculated using perturbation theory~\cite{Echevarria:2015byo,Li:2016ctv}.
However, when the soft function captures the soft-gluon effects at small $Q_\perp$, 
it is generically nonperturbative. 

Recently, using large 
momentum effective theory (LaMET)~\cite{Ji:2020ect,Ji:2013dva,Ji:2014gla}
a novel method has been proposed to extract the soft function from pion 
matrix elements (MEs)~\cite{Ji:2019sxk} that can be calculated in lattice QCD, 
enabling a solution of the difficult problem of nonperturbatively determining the soft function.
A first exploratory lattice QCD calculation was carried out
by the LPC Collaboration~\cite{Zhang:2020dbb}. 
However, a better understanding 
of the new method with a deeper examination of various systematic aspects involved in a 
lattice calculation is important in order to further establish the validity of the approach.

In this work, we perform a calculation of the soft function using
a different fermionic discretization, namely the twisted mass fermion.
We  demonstrate the validity of
the methodology proposed by Ref.~\cite{Ji:2019sxk} and determine the soft function, showing that to obtain the final results requires highly nontrivial steps including the following:
i) We apply an appropriate renormalization procedure~\cite{Orginos:2017kos,Izubuchi:2018srq,Gao:2020ito} to remove power and logarithmic divergences in the nonlocal operators; 
ii) we examine various pion MEs and find that some of the so-called
higher-twist (HT) contaminations are substantial and  can even flip the sign of the MEs.
By designing improved pion MEs to cancel the HT effects, we show that we can reliably 
obtain the soft function;  
iii)  we perform the calculation
at four different pion masses in order to examine the mass dependence of the soft function;
iv) we perform a detailed investigation of excited states;
and v) we  examine its convergence when the external momentum increases.

An important additional component of this work is the calculation of the Collins-Soper evolution kernel, where we find results that are in qualitative agreement with other lattice QCD calculations~\cite{Shanahan:2020zxr,Zhang:2020dbb,Schlemmer:2021aij}.

{\bf Theoretical framework} -- As proposed in Ref.~\cite{Ji:2019sxk}, the intrinsic, rapidity-independent soft function $S(b_\perp,\mu)$
depends on the transverse separation $b_\perp$ and the renormalization scale $\mu$.
Using LaMET, it can be extracted from the pion ME $F_{\Gamma}(b_\perp,P^z)$, which is defined in Euclidean spacetime as~\cite{Ji:2019sxk}
\begin{equation}
\label{eq:pion_matrix_element}
F_{\Gamma}(b_\perp,P^z)=\langle  \pi(-P^z) | \bar{u}\Gamma u (b_\perp)\, \bar{d}\Gamma d(0)|\pi(P^z)\rangle.
\end{equation}
Here, $P^z$ is a large momentum in the $z$ direction carried by the pion. Two current operators $\bar{u}\Gamma u$ and $\bar{d}\Gamma d$
are inserted at the same time slice, but with
a spatial separation $b_\perp$ that is perpendicular to the momentum direction.
To extract the leading-twist (LT) contribution, one can choose the Dirac matrices
as  $\Gamma=I,\gamma_5,\gamma_\perp$ or $\gamma_5 \gamma_\perp$.
$F_{\Gamma}(b_\perp,P^z)$ can be factorized into the quasi-TMD wave function (quasi-TMDWF) $\Phi$ and the intrinsic soft function $S(b_\perp,\mu)$~\cite{Ji:2019sxk,Ji:2020ect} at large $P^z$ through
\ba
    F_{\Gamma}(b_\perp,P^z)&\xlongequal[]{P^z\to \infty}&S(b_\perp,\mu) \int_{0}^{1} dx\,dx'\, H_{\Gamma}(x,x',P^z,\mu)
    \nn\\
    &&\hspace{0.5cm}\times{\Phi}^\dagger(x',b_\perp,-P^z){\Phi}(x,b_\perp,P^z),
\label{eq:factorization}
\ea
where $H_{\Gamma}(x,x',P^z,\mu)$ is the perturbative hard kernel.
The quasi-TMDWF $\Phi$ is defined as
\begin{equation}
\label{eq:quasi-TMDWF}
{\Phi}(x,b_\perp,\pm P^z)=\lim_{l\to\infty}\int\frac{d\xi}{2\pi}e^{ix \xi}\phi(\pm z,b_\perp,\pm l,\pm P^z)
\end{equation}
with $\xi=zP^z$. The wave function $\phi$ is given by
\begin{equation}
\phi(z,b_\perp,l,P^z)=\langle 0|O_{\phi}(t,z,b_\perp,l)|\pi(P^z)\rangle\, e^{E_\pi t}
\end{equation}
with $E_\pi=\sqrt{m_\pi^2+{P^z}^2}$. The operator $O_{\phi}$ is defined as
\begin{equation}
\hspace*{-0.2cm} O_{\phi}(t,z,b_\perp,l)\equiv\bar{u}(t,z/2, b_\perp) \Gamma_{\phi} W(z,b_\perp,l) d(t,-z/2,0).
\end{equation}
The quark fields $\bar{u},d$ and Wilson link $W$ entering $O_{\phi}$ are all located at the same time slice $t$.
 $W$ has a staple shape and goes through spatial sites
$(-z/2,0)\to(-l,0)\to(-l,b_\perp)\to(z/2,b_\perp)$. 
The Dirac matrix $\Gamma_{\phi}$ can be chosen as $\gamma_5\gamma_0$ or $\gamma_5\gamma_3$
so that $\Phi$ contains the
LT contribution. Here $\gamma_i$ ($i=0,1,2,3$) indicate the polarization direction $t,x,y,z$, respectively.

Up to $\mathcal{O}(\alpha_s)$ corrections,
the hard kernel takes a simple form, denoted here as $H_\Gamma^0$. It can be obtained from a Fierz identity that $H_{\Gamma}^0=1/(2N_c)$ for $\Gamma=I,\gamma_\perp,\gamma_5\gamma_\perp$ and
$-1/(2N_c)$ for $\Gamma=\gamma_5$ with $N_c=3$ the number of colors.
Using $H_{\Gamma}^0$ as an input, one can further simplify
the expression~(\ref{eq:factorization}) as
\begin{equation}
\label{eq:factorization_simple}
F_\Gamma(b_\perp,P^z)\xlongequal[\mathrm{LO\,\,kernel}]{P^z\rightarrow\infty}S(b_\perp,\mu)\, H^0_{\Gamma}\, |\phi(z=0,b_\perp,l=\infty,P^z)|^2.
\end{equation}
The soft function can be extracted by taking a ratio between $F_\Gamma$ and $H^0_{\Gamma}|\phi|^2$.
When using Eq.~(\ref{eq:factorization_simple}), one always fixes $z=0$. Thus, in the following context, the
variable $z$ is left out for simplicity.

{\em Lattice setup.} -- 
We use the gauge ensemble of $N_f=2+1+1$ clover-improved twisted mass fermions
generated by the Extended Twisted Mass Collaboration~\cite{Alexandrou:2018egz}.
In Eqs.~(\ref{eq:factorization}) and (\ref{eq:factorization_simple}), both $F_\Gamma$ and $\Phi$ contain
the structure information of the pion, which is expected to be
cancelled out at sufficiently large $P^z$, leaving the intrinsic soft function independent of either pion's structure
or its mass. To check the mass dependence, we use four valence quark masses,
corresponding to pion masses ranging 827 to 350 MeV. These valence quark masses together with other ensemble information are listed in
Table~\ref{tab:ensemble}.

\begin{table}
\centering

\setlength{\tabcolsep}{10 pt}
\begin{tabular}{cccccc}
	\hline
	\hline
	  $L/a$ & $T/a$&  $a$ (fm) &    $a\mu_{sea}$& $m^{\pi}_{sea}$ & $N_{meas}$ \\
     \hline
	 24 & 48 & 0.093  & 0.0053 & 350  & $126\times24$  \\
	 \hline
	
\end{tabular}
\setlength{\tabcolsep}{5.1 pt}
\begin{tabular}{cc|cc|cc|cc}
	\hline
$a\mu_{v0}$ & $m^{\pi}_{v0}$ & $a\mu_{v1}$ & $m^{\pi}_{v1}$ & $a\mu_{v2}$ &$m^{\pi}_{v2}$ &$a\mu_{v3}$ &$m^{\pi}_{v3}$ \\

0.0053 & 350 & 0.013 & 545 & 0.018 &640 & 0.03 & 827 \\
\hline
\end{tabular}
	\caption{Ensemble parameters used in this work. We list the spatial and temporal extents, $L/a$ and $T/a$,
	the lattice spacing $a$, the sea quark mass $\mu_{sea}$, the pion mass $m_{sea}^\pi$, the number of measurements $N_{meas}=N_{conf}\times(T/2)$, with
	$N_{conf}$ the number of configurations used,
	and four valence quark masses $\mu_{vi}$ for $i=0,1,2,3$ together with the associated pion masses $m_{vi}^\pi$. All pion masses are given in units of MeV.}
	\label{tab:ensemble}
\end{table}

The three-point correlation function for the pion ME is
\ba
&&C_\Gamma^{3pt}(b_\perp,P^z,t_s,t)=\frac{1}{L^3}\sum_{\vec{x}}e^{-2iP^zx_z}Z_\Gamma^2
\nn\\
&&\hspace{0.5cm}\langle O_\pi(t_s,-P^z)\,\bar{u}\Gamma u(t,\vec{x}+b_\perp)\,\bar{d}\Gamma d(t,\vec{x})\,O_\pi^\dagger(0,P^z)\rangle,
\ea
with $t_s$ the source-sink separation. The operators $\bar{u}\Gamma u$ and $\bar{d}\Gamma d$ are inserted at time slice $t$ and
constructed using the Coulomb-gauge-fixed-wall-source operators
\be
O_\pi(t,\vec{P})=\sum_{\vec{x},\vec{y}}\bar{u}(t,\vec{x})\gamma_5d(t,\vec{y})e^{-i\vec{P}\cdot\vec{y}},
\ee
which are known to
have a good overlap with the pion ground state. $Z_\Gamma$ is the renormalization factor to convert the bare lattice operator
$\bar{q}\Gamma q$ to the renormalized one in the $\overline{\mathrm{MS}}$ scheme.
The pion ME can be obtained from the connected part of three-point function at sufficiently large $t_s$ through
\begin{equation}
\label{eq:C_3pt}
	C^{3pt}_\Gamma\left(b_\perp,P^z,t_s,t\right)=\frac{|A_w(P^z)|^2}{(2E)^2} e^{-E_\pi t_s} F_{\Gamma}(b_\perp,P^z),
\end{equation}
where $A_w(P^z)=L^{-\frac{3}{2}}\langle\pi(P^z)|O_\pi^\dagger(0,P^z)|0\rangle$ is the overlap amplitude for the pion operator. According to parity, we have $A_w(P^z)=A_w(-P^z)$.

The correlation function for the quasi-TMDWF is constructed as
\begin{equation}
C_{\Gamma_\phi}^{wf}(b_\perp,l,P^z,t)=\frac{Z_\phi}{L^3}\sum_{\vec{x}}e^{-iP^zx_z}
\langle O_\phi(t,b_\perp,l)\,O_\pi^\dagger(0,P^z) \rangle,
\end{equation}
where $Z_\phi$ is the renormalization factor for the staple-shaped operator, which is found to be multiplicative~\cite{Ebert:2019tvc,Constantinou:2019vyb}.
We use $\Gamma_\phi=\gamma_5\gamma_0$ to avoid operator mixing for Wilson-type fermions
in the renormalization procedure~\cite{Constantinou:2019vyb}.
Stout smearing~\cite{Morningstar:2003gk} has been widely used in the lattice calculations involving nonlocal operators to reduce ultraviolet fluctuations.
Using up to 20 steps of smearing,  studies~\cite{Alexandrou:2018pbm,Alexandrou:2018eet,Alexandrou:2019lfo,Chai:2020nxw} demonstrate 
that the physics is not altered. Here, we apply 5 steps of smearing to construct 
the operator $O_\phi(t,b_\perp,l)$.
At large time separation $t$, one can extract the wave function $\phi$
via
\begin{equation}
\label{eq:C_TMDWF}
C^{wf}_{\Gamma_\phi}(b_\perp,l,P^z,t)=\frac{A_w(P^z)}{2E_\pi} e^{-E_\pi t}\phi(b_\perp,l,P^z).
\end{equation}

Combining Eqs.~(\ref{eq:C_3pt}) and (\ref{eq:C_TMDWF}), the intrinsic soft function defined in Eq.~(\ref{eq:factorization_simple}) can be obtained through
\begin{equation}
S(b_\perp)=\lim_{l\to\infty}\lim_{t_s\to\infty}\frac{C^{3pt}_{\Gamma}\left(b_\perp,P^z,t_s,t\right)}{H_{\Gamma}^0\,\left|C^{wf}_{\Gamma_\phi}(b_\perp,l,P^z,\frac{t_s}{2})\right|^2}.
\end{equation}
The lattice data show that $C^{wf}_{\Gamma_\phi}$ carries a small but nonvanishing imaginary part.
In the determination of $S(b_\perp)$, we take into account the contributions
from both the real and imaginary part of the wave function.

To examine the convergence of the lattice results at large momentum, we utilize 8 momenta with $P^z=\pm n(2\pi/L)$ ($n=3,4,5,6$), corresponding to a range from $\pm 1.7$ to $\pm3.3$ GeV.
Given each $P^z$, we average the transition modes $\pi(P^z)\to\pi(-P^z)$ and $\pi(-P^z)\to\pi(P^z)$ and obtain a 15\%-20\% reduction in the statistical error.
For each momentum, we place the wall-source operator at every two time slices, which allows us to perform a time translation average for
both $C_\Gamma^{3pt}$ and $C^{wf}_{\Gamma_\phi}$. This helps to reduce the uncertainty of the soft function by nearly a factor of $\sqrt{T/2}$.

{\em Renormalization.} -- In our past calculation of the nucleon and Delta
quasi-PDFs~\cite{Alexandrou:2017huk,Chai:2020nxw},
we have utilized the regularization-independent momentum-subtraction (RI-MOM) scheme~\cite{Martinelli:1994ty} developed for nonlocal operators~\cite{Constantinou:2017sej,Alexandrou:2017huk}. The
RI-MOM renormalized correlator is defined as $C_{\Gamma_\phi}^{wf,RI}=C_{\Gamma_\phi}^{wf,b}Z_\phi^{RI}$,
with the renormalization factor $Z_\phi^{RI}$ extracted by evaluating the amputated vertex functions with quark external
states. This renormalization factor cancels the power
and logarithmic divergences up to some systematic effects, such as discretization and HT effects.
When using the staple-shaped operator, the systematic effects, which enter in the renormalization procedure, become more complicated.
In our calculation, we use the ratio scheme~\cite{Orginos:2017kos,Izubuchi:2018srq,Gao:2020ito} instead,
which has been proposed to replace the quark-state MEs in RI-MOM by the corresponding hadronic ones
for a better control of systematics, such as discretization and HT effects.
Here, we adopt the ratio scheme and construct the renormalized
correlator as
\be
\label{eq:ratio_scheme}
C_{\Gamma_\phi}^{wf,r}(b_\perp,l,P^z,t)=\frac{C_{\Gamma_\phi}^{wf,b}(b_\perp,l,P^z,t)}{C_{\Gamma_\phi}^{wf,b}(b_\perp,l,0,t)}C_{\Gamma_\phi}^{wf,\overline{\mathrm{MS}}}(0,0,0,t),
\ee
where the bare correlators $C_{\Gamma_\phi}^{wf,b}(b_\perp,l,P^z,t)$ and $C_{\Gamma_\phi}^{wf,b}(b_\perp,l,0,t)$ contain the same operator $O_\phi(t,b_\perp,l)$ and only differ by $P^z$. Thus, one can expect that a clean cancellation of UV divergences and other systematics can be achieved using the ratio scheme. 
Note that the renormalization has already been accomplished when taking the ratio $C_{\Gamma_\phi}^{wf,b}(b_\perp,l,P^z,t)/C_{\Gamma_\phi}^{wf,b}(b_\perp,l,0,t)$.
The coefficient $C_{\Gamma_\phi}^{wf,\overline{\mathrm{MS}}}(0,0,0,t)$ is introduced to restore the correct normalization at $b_\perp\to0$.
The conversion from the ratio scheme to the $\overline{\mathrm{MS}}$ scheme would require a perturbative calculation of the $\alpha_s\ln(b_\perp\mu)$ corrections. 
Up to these corrections one can treat $C_{\Gamma_\phi}^{wf,r}(b_\perp,l,P^z,t)$ and $C_{\Gamma_\phi}^{wf,\overline{\mathrm{MS}}}(b_\perp,l,P^z,t)$ as the same.

The renormalization for the local current operator $\bar{u}\Gamma u$ or $\bar{d}\Gamma d$ in $C_\Gamma^{3pt}$ is straightforward.
We find $Z_S^{\overline{\mathrm{MS}}}(2\,{\rm GeV})=0.641(3)$, $Z_P^{\overline{\mathrm{MS}}}(2\,{\rm GeV})=0.475(4)$ and $Z_V=0.712(2)$, $Z_A=0.753(3)$. Note that $Z_V$ and $Z_A$ are scheme and scale independent.
These are calculated on dedicated $N_f=4$ ensembles with the same lattice action and spacing as the $N_f=2+1+1$ ensemble used
for the MEs.
The definitions of $Z_V$, $Z_A$, $Z_S$, and $Z_P$ follow the convention of Ref.~\cite{Alexandrou:2015sea}.
Note that when the two operators $\bar{u}\Gamma u(b_\perp)$ and $\bar{d}\Gamma d(0)$ approach each other, a contact term appears and
additional renormalization is required to match two bilinear quark operators to a local four-quark operator. 
Since the four-quark operators do not mix with any lower dimensional operators, we expect that additional renormalization effects
are not large. It has been shown that  renormalization factor for the four-quark operator only differs from that of the product of 
two local operators by 20\%~\cite{Christ:2012se,RBC:2010qam}. 
Such effects require further investigation but are not expected to alter the conclusions of this work.

{\em Systematic effects.} -- In Eq.~(\ref{eq:quasi-TMDWF}), the quasi-TMDWF is defined at an infinitely-large length of the Wilson line $l$.
In a realistic lattice calculation, $l$ is truncated by a finite lattice size.
At sufficiently large $l$ , we find that the lattice results of $|C_{\Gamma_\phi}^{wf,r}|$ converge 
and yield a plateau for the region of $l\gtrsim0.8$ fm (see the Supplemental Material).
Thus, fits to a constant lead to good $\chi^2/{\mathrm{dof}}$ and provide the results of $|C_{\Gamma_\phi}^{wf,r}|$ at $l\rightarrow\infty$.

To extract reliably the pion ME from $|C_{\Gamma_\phi}^{wf,r}|$, the excited-state contamination is another systematic effect to be controlled.
We calculate the correlation functions at $t_s/a=6,8,10,12$ and use these data to perform a two-state fit. 
The lattice results are shown after removing the excited-state contamination.

After taking the extrapolation of $l\to\infty$ and examining the ground-state saturation at sufficiently large $t$, we use the simplified notation $C^{wf}_{\Gamma_\phi}(b_\perp, P^z)$
to replace $C^{wf}_{\Gamma_\phi}(b_\perp, l, P^z,t)$. To reveal the systematic effects more clearly,
all figures presented in this work are compiled using the most
precise lattice data at $m_\pi=827$ MeV, unless specified otherwise.

\begin{figure}
	\includegraphics[width=0.48\textwidth,angle=0]{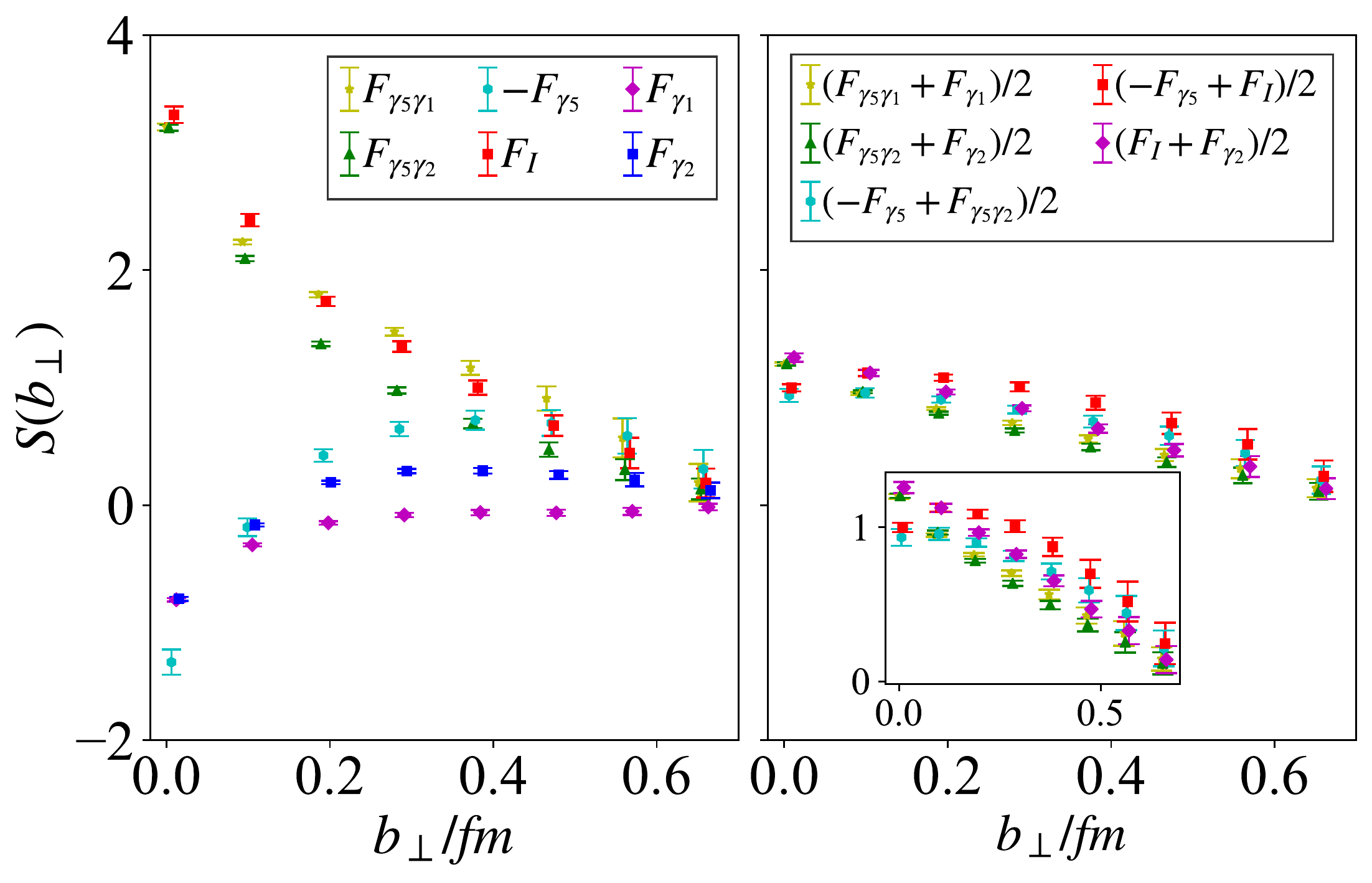}
	\caption{The intrinsic soft function $S(b_\perp)$ as a function of transverse separation $b_\perp$ at $P^z=6\frac{2\pi}{L}\approx3.3$ GeV and $m_\pi=827$ MeV.
	On the left panel, $S(b_\perp)$ are compiled using the pion MEs $F_\Gamma$ as inputs, with $\Gamma=I,\gamma_5,\gamma_\perp,\gamma_5\gamma_\perp$. For $\gamma_\perp$, there are two choices: $\gamma_1$ parallel to $b_\perp$ and $\gamma_2$ perpendicular to $b_\perp$. On the right panel, $S(b_\perp)$ are compiled using the improved
pion MEs, where the large HT contamination has been canceled significantly and the results show much better consistency.}
	\label{fig:higher_twist_contamination}
\end{figure}

{\bf Extraction of LT contribution} -- According to the proposal of Ref.~\cite{Ji:2019sxk}, at $P^z\to\infty$, the same intrinsic soft function can be extracted from
various pion MEs $F_\Gamma$ as far as $F_\Gamma$ contain the LT contribution.
In this work, we make a complete investigation of the $\Gamma$ dependence of the soft function. Fig.~\ref{fig:higher_twist_contamination} (left) illustrates that the results of the soft function are significantly different
when using various $F_\Gamma$ as inputs. Some results even carry the opposite sign.

To resolve this puzzle, we check the factorization in the LO perturbation theory and find at finite $P^z$
\ba
\label{eq:pion_matrix_element_decomposition}
F_\Gamma(b_\perp,P^z)&=&S(b_\perp)\,H_\Gamma^0\,|\phi(b_\perp,l,P^z)|^2
\nn\\[1ex]
&+&\sum_{\Gamma'\neq\gamma_5\gamma_0,\gamma_5\gamma_3} 
\hspace*{-0.25cm}
S_{\Gamma'}(b_\perp)\,H_{\Gamma\Gamma'}^0\,|\phi_{\Gamma'}(b_\perp,l,P^z)|^2+\cdots,
\nn\\
\ea
where the factor $H_{\Gamma\Gamma'}^0$ arises from Fierz rearrangement through
\ba
\hspace*{-0.65cm}
\bar{u}\Gamma u(b_\perp)\bar{d}\Gamma d(0)=\sum_{\Gamma'} H_{\Gamma\Gamma'}^0
\bar{u}(b_\perp)\Gamma' d(0) \bar{d}(0)\Gamma'u(b_\perp)
\ea
with $H_{\Gamma\Gamma'}^0=\frac{1}{16N_c}\operatorname{Tr}(\Gamma\Gamma'\Gamma\Gamma')$.
The LT contribution carries a factor of $H_\Gamma^0$,
which is the summation of $H_{\Gamma\Gamma'}^0$ with $\Gamma'=\gamma_5\gamma_0$ and $\gamma_5\gamma_3$.
HT contributions enter in the second term of Eq.~(\ref{eq:pion_matrix_element_decomposition}) with
the wave function $\phi_{\Gamma'}(b_\perp,l,P^z)=\langle0|\bar{u}(b_\perp)\Gamma'W(b_\perp,l)d(0)|\pi(P^z)\rangle$.
Although HT contributions are expected to be much smaller than the LT one at sufficiently large momentum,
in a realistic lattice calculation, where the typical size of $P^z$ is a few GeV, the contamination from HT may be significant.
We find that the lattice result of $\phi_{\Gamma'}$ for $\Gamma'=\gamma_5$ is even larger than the LT $\phi$.
Such large HT contamination explains why some $F_\Gamma(b_\perp,P^z)$ carry the opposite sign,
as observed in Fig.~\ref{fig:higher_twist_contamination}. Here we focus on the largest power corrections associated with $\phi_{\Gamma'}$. Any residual
corrections are represented by the ellipsis in Eq.~(\ref{eq:pion_matrix_element_decomposition}).

Note that in Fig.~\ref{fig:higher_twist_contamination}, results at the largest momentum $P^z=6\frac{2\pi}{L}\approx3.3$ GeV are presented. When $P^z$ decreases,
the situation becomes even worse. This is not surprising, as LT contributions are enhanced at large $P^z$.
Considering the fact that the $P^z$ values accessible on the lattice are quite limited,
we draw the conclusion that it is essential to remove the HT effects in the calculation of the soft function.
Here we take two steps.
\begin{itemize}
\item First, we calculate $\phi_{\Gamma'}$ with various $\Gamma'$ and then pick up all $\phi_{\Gamma'}$ with relatively large size.
It leads to four $\phi_{\Gamma'}$ with $\Gamma'=\gamma_5,\sigma_{02},\sigma_{12},\sigma_{23}$. 
\item Second, we define improved pion MEs as $\sum_{\Gamma} c_\Gamma F_\Gamma(b_\perp,P^z)$, where
the coefficients $c_\Gamma$ ($\Gamma=I,\gamma_5,\gamma_\perp,\gamma_5\gamma_\perp$) are chosen appropriately to
cancel contributions from $\phi_{\Gamma'}$ ($\Gamma'=\gamma_5,\sigma_{02},\sigma_{12},\sigma_{23}$).
\end{itemize}

Following the above steps, we finally obtain five improved
pion MEs as a simple combination of two $F_\Gamma(b_\perp)$, namely
\ba
&&\frac{1}{2}(F_{\gamma_5\gamma_1}+F_{\gamma_1}), \quad \frac{1}{2}(F_{\gamma_5\gamma_2}+F_{\gamma_2}),\quad \frac{1}{2}(-F_{\gamma_5}+F_{\gamma_5 \gamma_2}),
\nn\\
&&\frac{1}{2}(-F_{\gamma_5}+F_{I}),\quad\frac{1}{2}(F_{I}+F_{\gamma_2}).
\label{type}
\ea
Fig.~\ref{fig:higher_twist_contamination} (right) shows the soft function compiled using the five improved pion MEs. By canceling the dominant HT effects, the results become much more consistent. Residual deviations serve as measure of important systematic effects
to be controlled in future studies.

{\em Results of the soft function.} --
After checking the consistency among the various improved pion MEs, we use the choice of $\frac{1}{2}\left(F_{\gamma_5\gamma_1}+F_{\gamma_1}\right)$
as an example to present the results of $S(b_\perp)$ for various momenta $P^z$ and pion masses $m_{vi}^\pi$.

\begin{figure}
	\includegraphics[width=0.45\textwidth,angle=0]{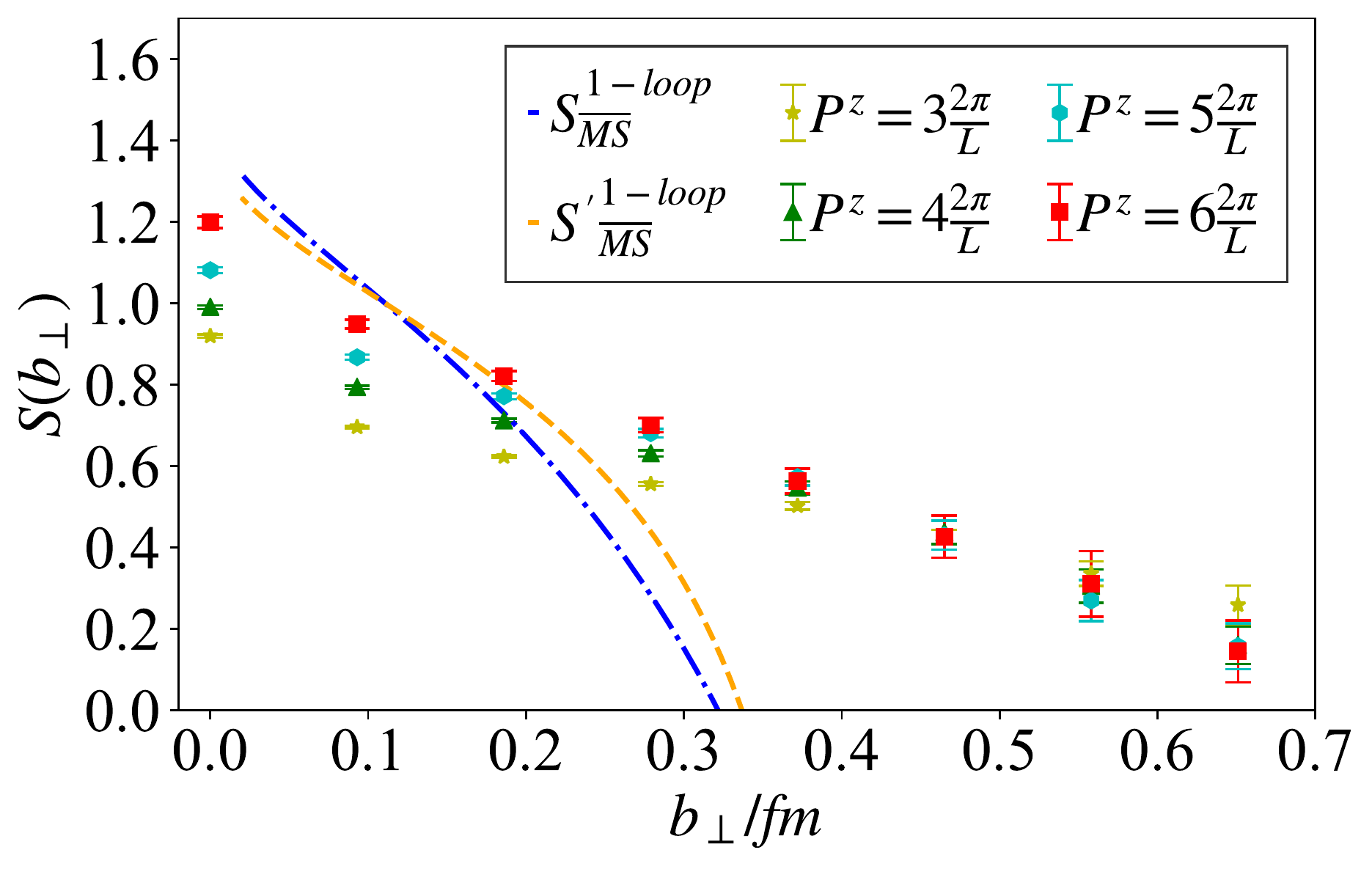}
	\caption{The lattice results of $S(b_\perp)$ for various momenta at $m_\pi=827$ MeV, together with the one-loop
	perturbative result $S^{\rm 1-loop}_{\overline{\rm MS}}$ and its variant $S'^{\rm 1-loop}_{\overline{\rm MS}}$ with $\alpha_s$ including up to 4 loops. The scale $\mu$ in Eq.~(\ref{eq:soft_function_one_loop}) is set as $\mu=2$ GeV.}
	\label{fig:soft_function}
\end{figure}

In Fig.~\ref{fig:soft_function}, $S(b_\perp,P^z)$ is shown together with the one-loop perturbative curve~\cite{hep-ph/0404183},
\begin{equation}
\label{eq:soft_function_one_loop}
S_{\overline{\rm MS} }(b_\perp,\mu)=1-\frac{\alpha_s C_F}{\pi} \ln\frac{\mu^2 b_{\perp}^2}{4 e^{-2\gamma_E}} +\mathcal{O}(\alpha_s^2),
\end{equation}
where one-loop and four-loop values of $\alpha_s$ are used at the physically most relevant scale of $S(b_\perp)$, i.e.\ $1/b_\perp$. The scale $\mu$ is set as $\mu=2$ GeV.
We note that the lattice results agree qualitatively with the perturbative function at around $b_\perp\sim 0.2$~fm, particularly at the largest boost and when higher-order effects are partially included via $\alpha_s$.
At larger $b_\perp$, nonperturbative features start to set in and the decay of $S(b_\perp)$ is slower than the perturbative prediction.
It is also noteworthy that the convergence of the lattice results in $P^z$ clearly increases with $b_\perp$ -- the results from the two largest $P^z$ are compatible for $b_\perp\gtrsim0.2$ fm, while smaller transverse separations will need yet larger boosts to establish convergence.

\begin{figure}
	\includegraphics[width=0.45\textwidth,angle=0]{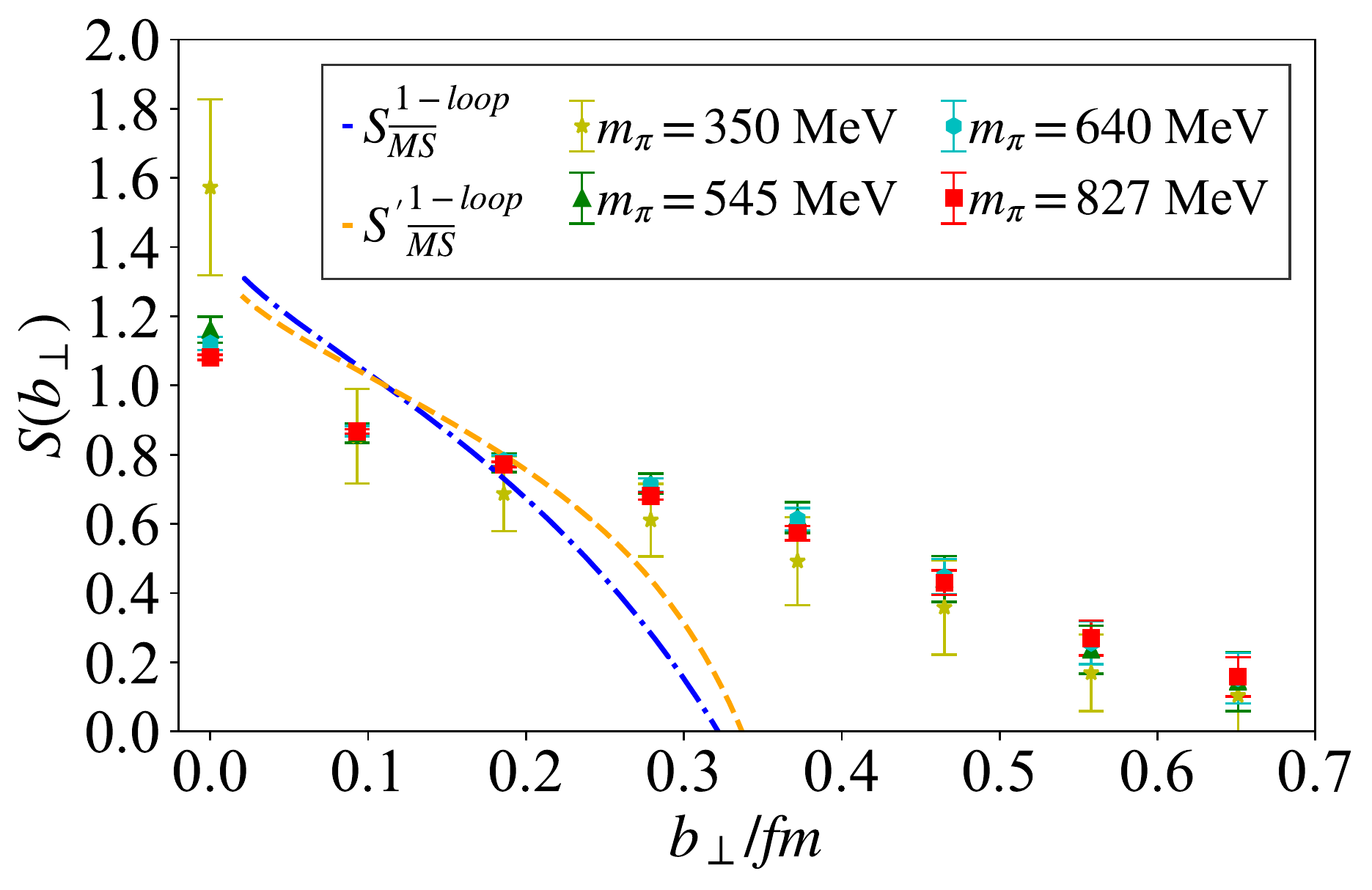}
	\caption{The intrinsic soft function $S(b_\perp)$ for the pion masses ranging from
827 to 350 MeV. Here, we show results calculated at $P^z=5\frac{2\pi}{L}$ as an example.
}
	\label{fig:pion_mass_dep}
\end{figure}

In Fig.~\ref{fig:pion_mass_dep}, we examine the pion mass dependence of the soft function. Although $S(b_\perp)$ is extracted from pion MEs which depend on the detailed process of $\pi(P^z)\to\pi(-P^z)$, the factorization allows us to cancel this process dependence. Performing the calculation at four pion masses, we find that the lattice results are generally consistent within statistical errors, although a small systematic increase is found when decreasing $m_\pi$.
Within current errors, this observation is consistent with the expectation from factorization theory~\cite{Ji:2019sxk} that the soft function should not depend on the detailed hadronic information from the initial or final state.

{\em Results for the Collins-Soper kernel.} -- 
The Collins-Soper kernel $K(b_\perp,\mu)$ 
governs the rapidity evolution of the TMDPFs. In LaMET, the quasi-TMDPDF is 
factorized into the light-cone TMDPDF and a $K(b_\perp,\mu)\ln(\zeta^z/\zeta)$  
factor, where $\zeta^z=2(xP^z)^2$, with $P^z$ playing the role of the rapidity, 
while $\zeta$ is the light-cone counterpart of $\zeta^z$~\cite{Ebert:2018gzl}. 
Thus, by taking the ratio of quasi-TMDPDFs at different values of $P^z$, one can extract 
$K(b_\perp,\mu)$. This ratio can also be expressed in terms of the quasi-TMDWFs~\cite{Zhang:2020dbb} as 
\ba
\label{eq:new_kernel}
\hspace{-0.7cm}K(b_\perp,\mu)&=&\lim_{l\to\infty}\frac{1}{\ln(P_1^z/P_2^z)}
\ln\left|\frac{\phi(b_\perp,l,P^z_1)/E_1}{\phi(b_\perp,l,P^z_2)/E_2}\right|
\nn\\
&=&\frac{1}{\ln(P_1^z/P_2^z)}\ln\left|\frac{C^{wf}_{\Gamma_\phi}(b_\perp,P^z_1)}{C^{wf}_{\Gamma_\phi}(b_\perp,P^z_2)} \frac{C^{wf}_{\Gamma_\phi}(0,P^z_2)}{C^{wf}_{\Gamma_\phi}(0,P^z_1)}\right|.
\ea

\begin{figure}[H]
	\includegraphics[width=0.45\textwidth,angle=0]{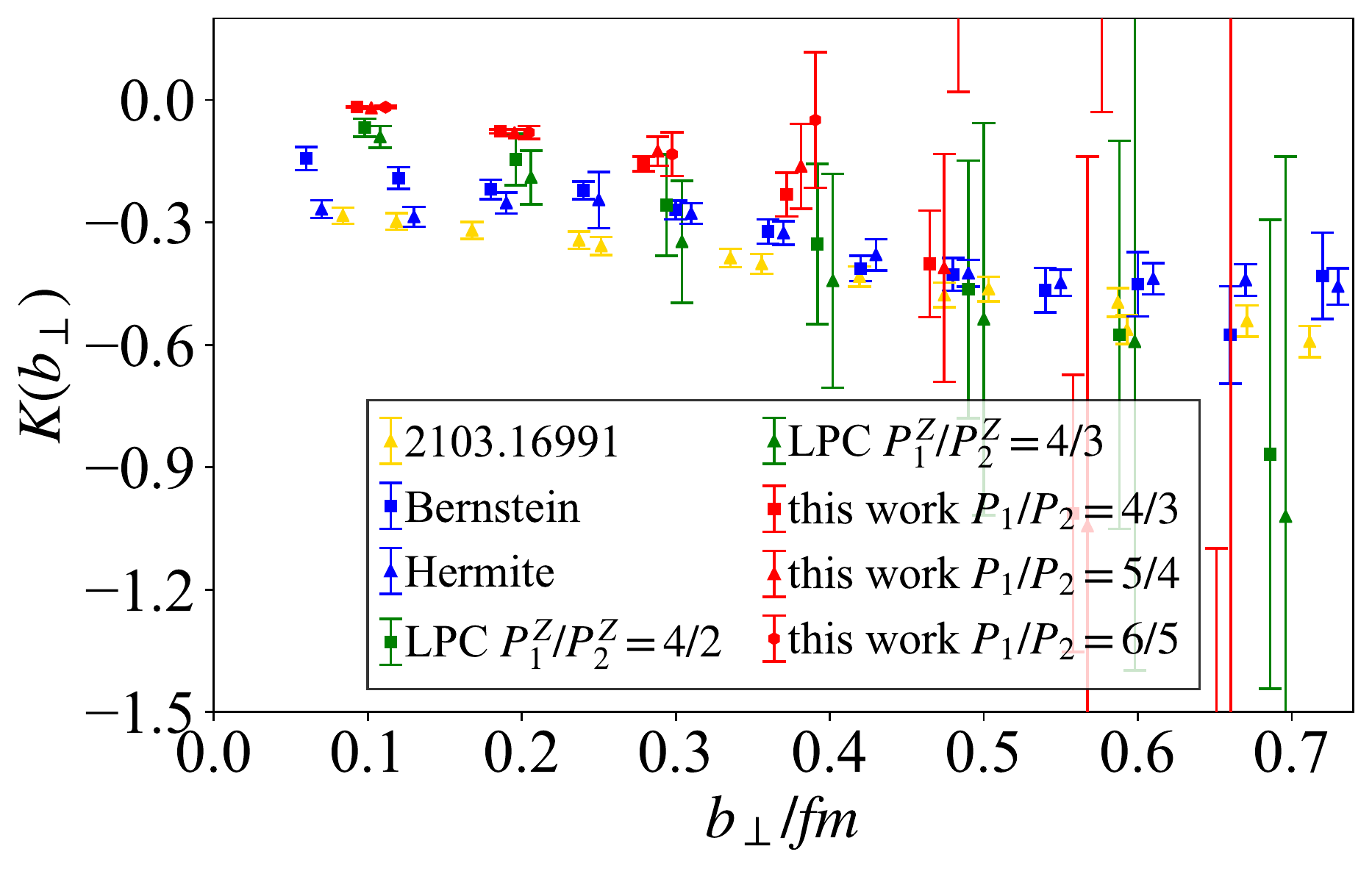}
	\caption{The lattice results for the Collins-Soper kernel $K(b_\perp,\mu)$ from various calculations, described by 
yellow~\cite{Schlemmer:2021aij}, blue~\cite{Shanahan:2020zxr}, green~\cite{Zhang:2020dbb} and red. Results from the same calculation are shifted horizontally to make an easier comparison. For this work, the setup at $m_\pi=827$ MeV is used.}
	\label{fig:CS_kernel}
\end{figure}

In Fig.~\ref{fig:CS_kernel}, the lattice results of $K(b_\perp,\mu)$ from this work are shown together with
data from other calculations.
The results exhibit similar dependence on $b_\perp$ with some discrepancies, which indicate unquantified systematics.
Both the LPC results and ours are calculated using the quasi-TMDWFs as inputs. Thus, it is not surprising that these results are in better agreement.

{\em Conclusion.} --
Within the framework of lattice QCD we calculate the intrinsic soft function introducing a number of crucial steps that enable its reliable extraction.
Our work adds evidence that the methodology proposed in Ref.~\cite{Ji:2019sxk} is indeed suitable for the determination of these quantities. There is room for further improvements.
For example, only the LO perturbative hard kernel is used in this calculation and future work needs to examine higher-order corrections.
On the lattice side, several sources of systematics need to be addressed, including e.g.\ cutoff effects and further investigation of quark mass dependence towards the physical one.
Nevertheless, this methodology coupled with the improvements introduced in this work, requiring synergy of perturbative and lattice QCD, is shown to be very promising and can provide important first-principle insights into TMD hadron structure.

\begin{acknowledgements}
We thank Lu-Chang Jin, Yi-Zhuang Liu, Yu-Sheng Liu, Yan-Qing Ma, Wei Wang, Yi-Bo Yang, Qi-An Zhang and Yong Zhao for valuable discussions.
We thank Maximilian Schlemmer, Qi-An Zhang and Yong Zhao for providing their data of the Collins-Soper kernel.
X.F. and S.C.X. are supported in part by NSFC of China under Grants 
No. 12125501, No. 12141501, and No. 11775002
and National Key Research and Development Program of China under Contracts No. 2020YFA0406400.
X.F. and C.L. are supported in part by NSFC of China under Grant No. 12070131001.
Y.L. and C.L. are supported in part by CAS Interdisciplinary Innovation Team and NSFC of China under Grant No. 11935017.
F.S. was funded  by the NSFC and the Deutsche Forschungsgemeinschaft (DFG, German Research
Foundation) through the funds provided to the Sino-German Collaborative
Research Center TRR110 “Symmetries and the Emergence of Structure in QCD”
(NSFC Grant No. 12070131001, DFG Project-ID 196253076 - TRR 110).
K.C. is supported by the National Science Centre (Poland) grant SONATA BIS No. 2016/22/E/ST2/00013.
K.H. is financially supported by the Cyprus Research and Innovation foundation under contract number POST-DOC/0718/0100.
M.C. and A.S. acknowledge financial support by the U.S. Department of Energy, Office of Nuclear Physics, Early Career Award under Grant No.\ DE-SC0020405.
J.T. acknowledges support from project NextQCD, co-funded by the European Regional Development Fund and the Republic of
Cyprus through the Research and Innovation Foundation (EXCELLENCE/0918/0129).
The calculation was carried out on TianHe-3 (prototype) at Chinese National Supercomputer Center in Tianjin.
This work also used computational resources from the John von Neumann-Institute for Computing on the Juwels booster system at the research center in Juelich, under the project with id ECY00
and on the Cyclone machine of the Cyprus Institute under project ID pro21a106.

\end{acknowledgements}

\bibliography{reference}

\bibliographystyle{h-physrev}

\clearpage

\setcounter{page}{1}
\renewcommand{\thepage}{Supplementary Information -- S\arabic{page}}
\setcounter{table}{0}
\renewcommand{\thetable}{S\,\Roman{table}}
\setcounter{equation}{0}
\renewcommand{\theequation}{S\,\arabic{equation}}

\section{Supplementary Material}
In this section, we expand on a selection of technical details and add
results to facilitate cross-checks of different calculations of
the soft function.

\begin{figure}[htb]
	\includegraphics[width=0.45\textwidth,angle=0]{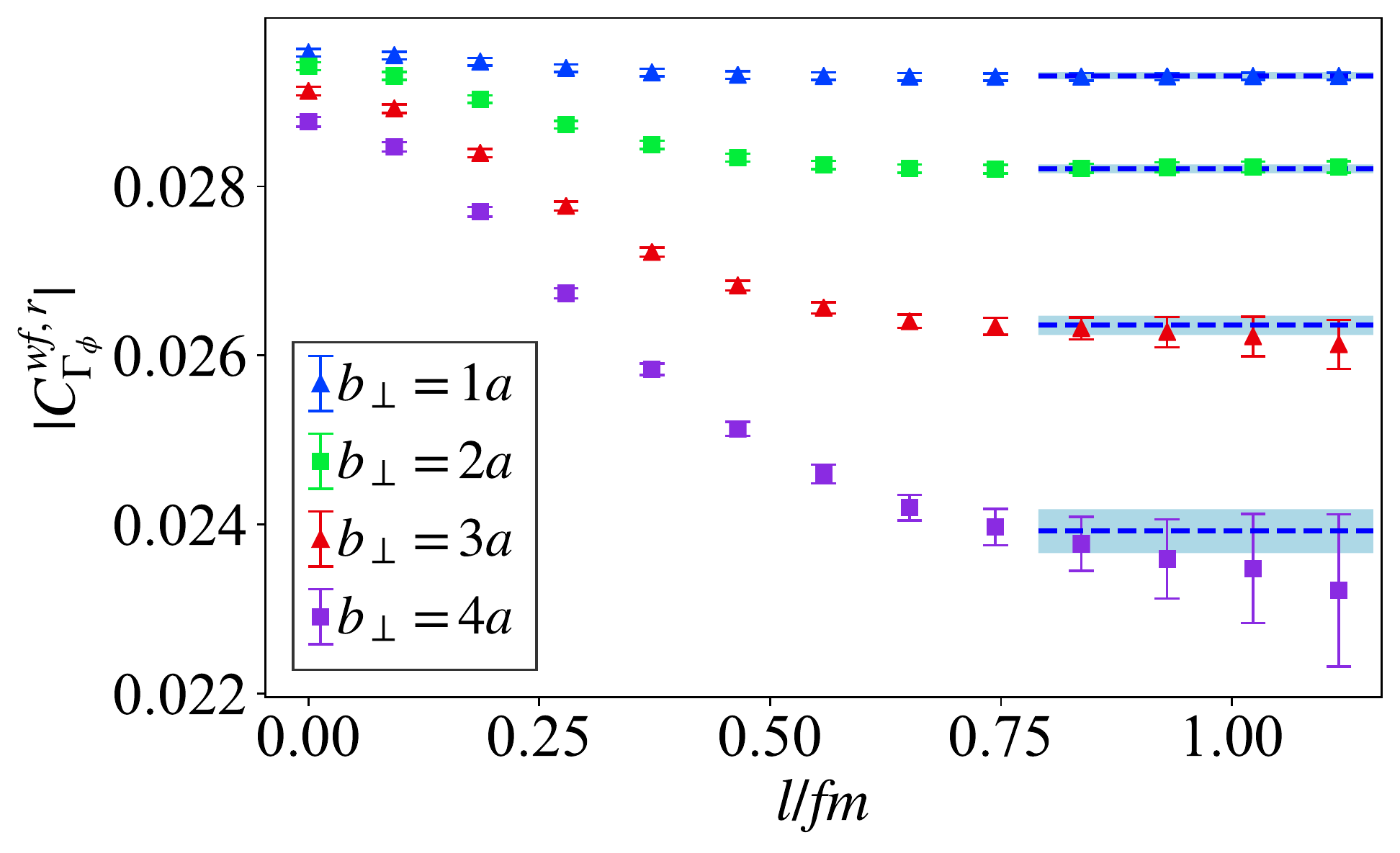}
	\includegraphics[width=0.45\textwidth,angle=0]{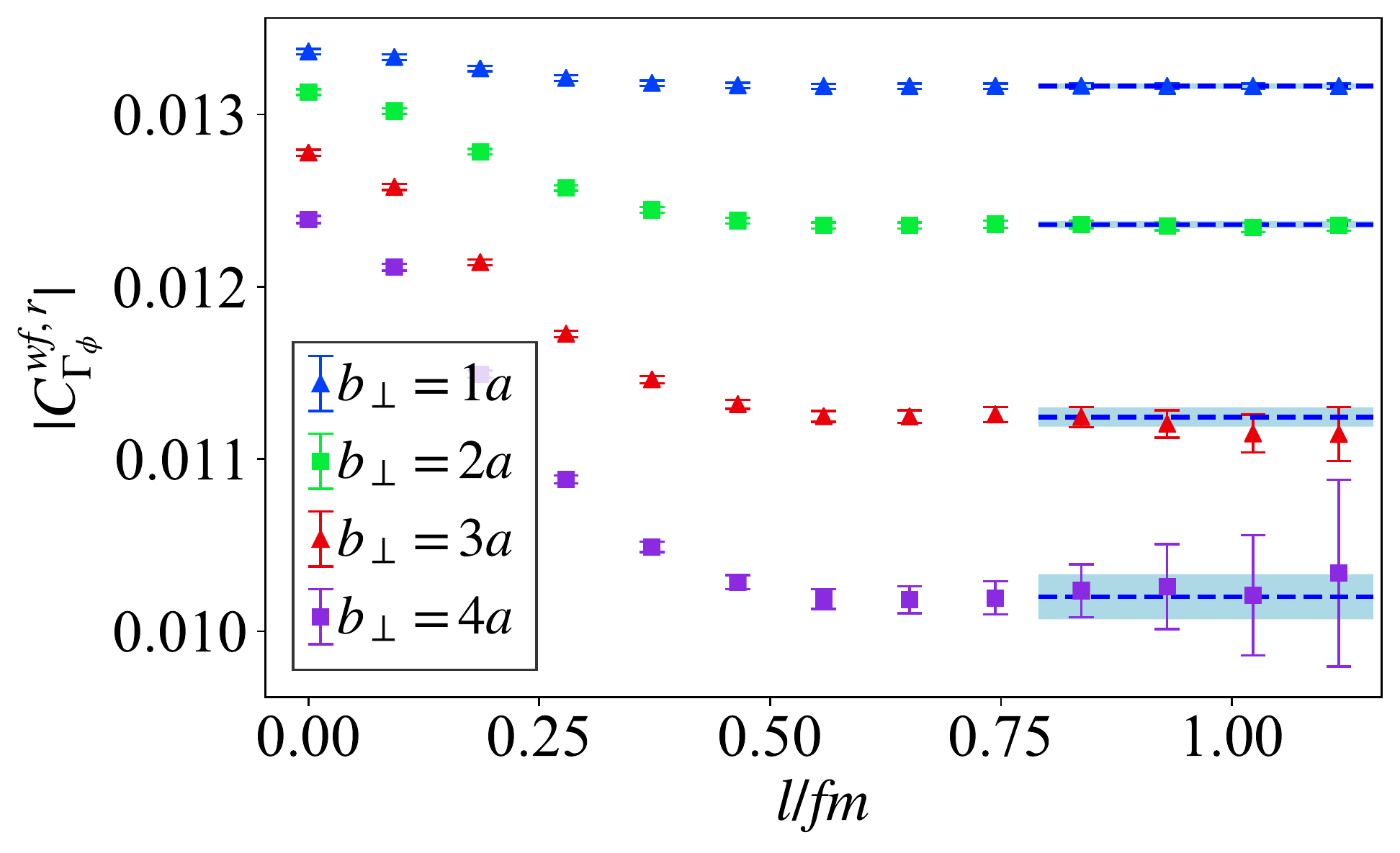}
	\caption{The $l$ dependence of the renormalized correlation function $|C_{\Gamma_{\phi}}^{wf,r}|$ at $t/a=5$ and $P^z=3 \frac{2\pi}{L}$ (top) and $5\frac{2\pi}{L}$ (bottom).
	Results at four different $b_\perp$ are shown. The $\chi^2/\mathrm{dof}$, which describes the quality of the correlated fit, is listed as
\{1.1,0.3,0.4,0.6\} for the case of $P^z=3 \frac{2\pi}{L}$ and \{0.8,1.4,0.4,0.3\}
for $5\frac{2\pi}{L}$. Here we use the results at $m_\pi=827$ MeV as an example. The same plateau range is found for other pion masses.}
	\label{fig:L_dependence}
\end{figure}

{\bf \boldmath Extrapolation of $l\to\infty$} --
Although in the lattice QCD calculation the length of the Wilson link $l$ is not allowed to be larger than half of the lattice size $L/2$,
it is straightforward to explore the limit of $l\to\infty$ if the renormalized correlation function $|C_{\Gamma_\phi}^{wf,r}|$ has a plateau at large $l$.
In Fig.~\ref{fig:L_dependence}, we show two examples with the external momentum $P^z=3\frac{2\pi}{L}$ and $5\frac{2\pi}{L}$. In both cases, the plateau appears
when $l\ge 0.84$ fm.
Using a correlated fit to the constant and extrapolating to the $l\to\infty$ limit, we
finally obtain the results of $|C_{\Gamma_\phi}^{wf,r}|$ at $l=\infty$.

\begin{figure}[htb]
	\includegraphics[width=0.45\textwidth,angle=0]{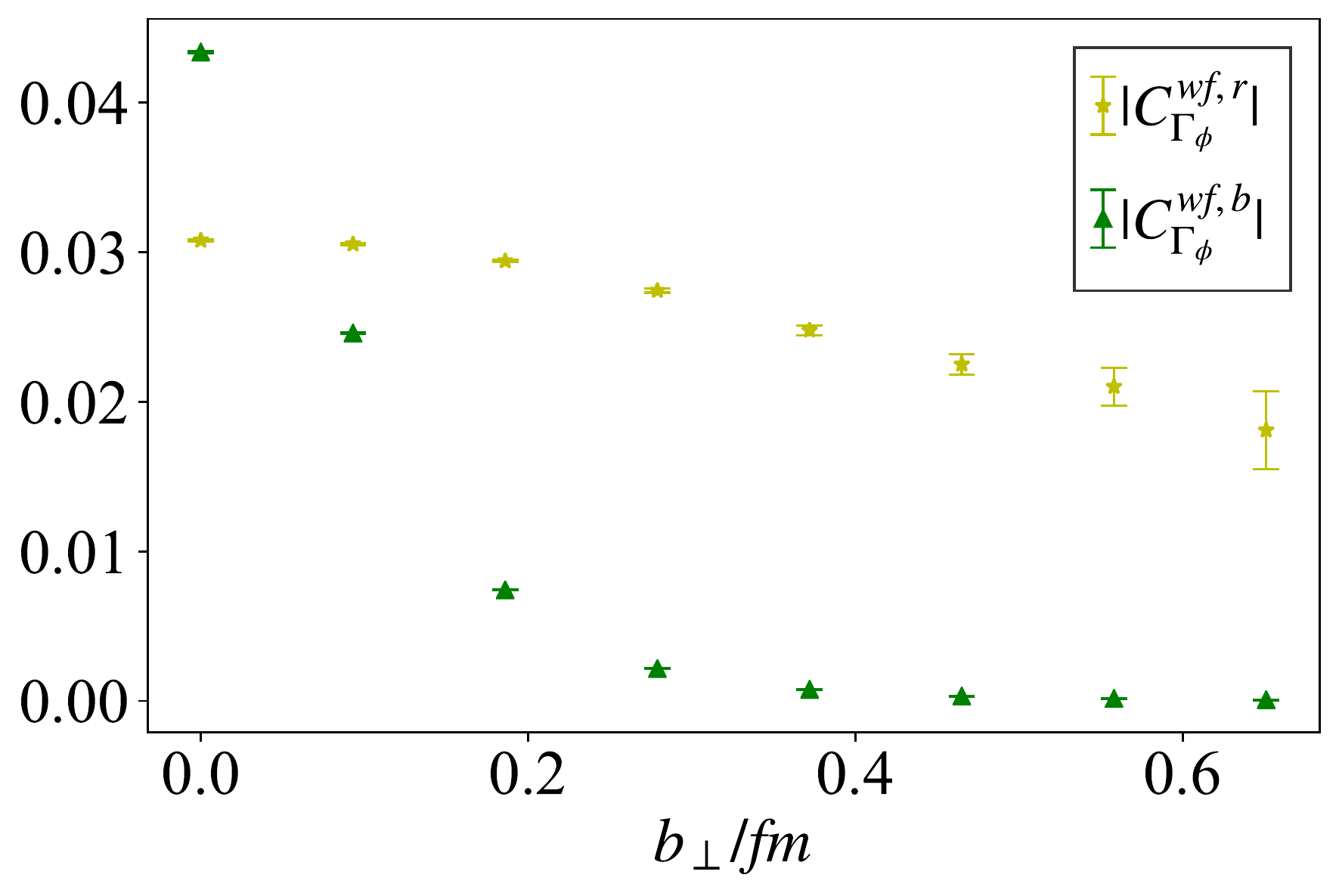}
	\caption{A comparison between the renormalized correlation function $C_{\Gamma_\phi}^{wf,r}$ constructed using Eq.~(\ref{eq:ratio_scheme})
	and the bare one $C_{\Gamma_\phi}^{wf,b}$.
	The correlation functions at $t/a=5$, $P^z=3 \frac{2\pi}{L}$ and $m_\pi=827$ MeV are shown as a function of $b_\perp$.}
	\label{fig:Bare_VS_Renorm}
\end{figure}

In Fig.~\ref{fig:Bare_VS_Renorm}, we show a comparison between the renormalized and bare correlation functions.
The very different $b_\perp$ dependence suggests that it is crucial to apply the renormalization procedures to remove the ultraviolet divergence.

\begin{figure}[htb]
	\includegraphics[width=0.45\textwidth,angle=0]{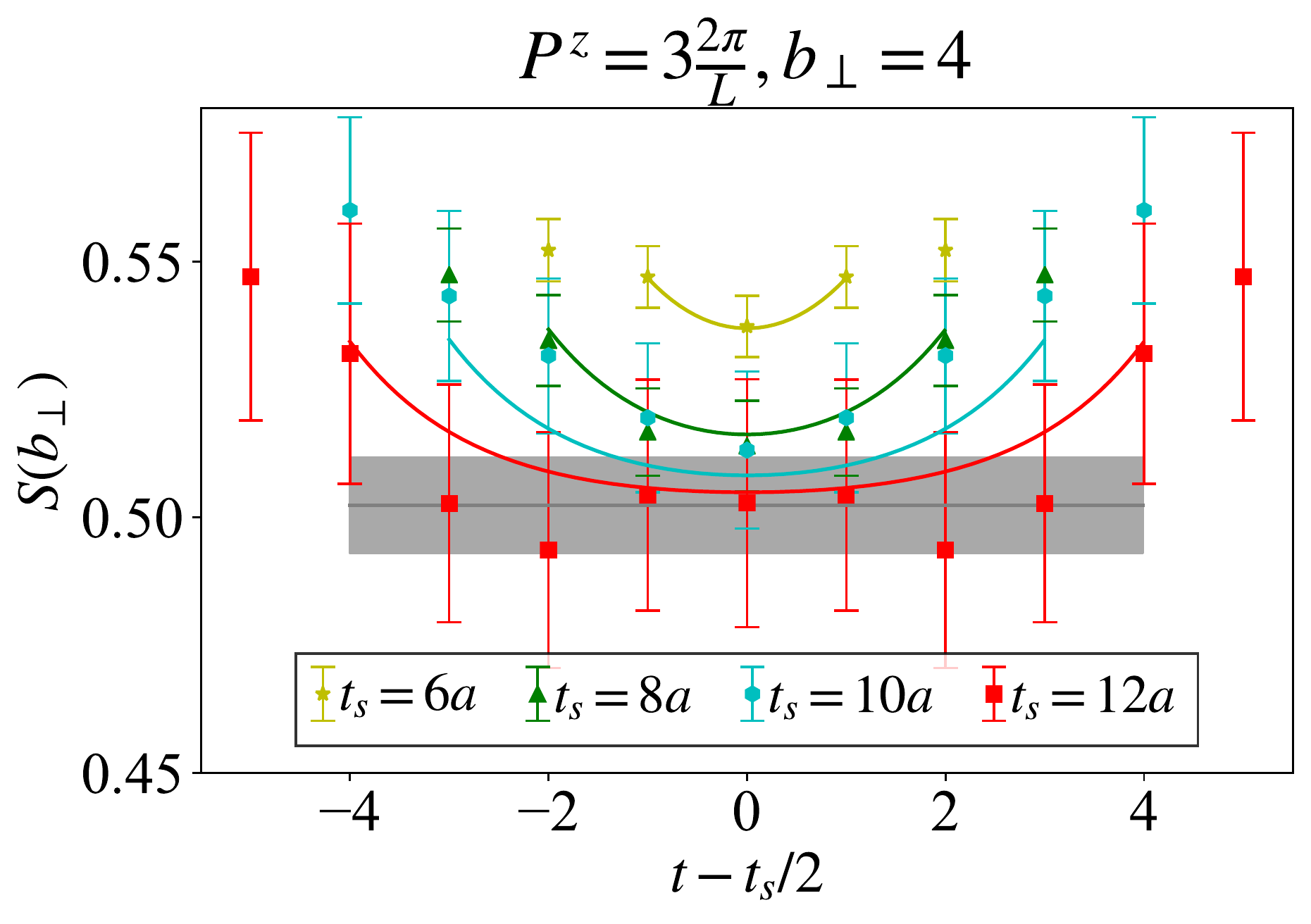}
	\caption{A comparison of soft function with $\{P^z,b_\perp,m_\pi\}=\{3\frac{2\pi}{L},4a,\mbox{827 MeV}\}$ 
	for different source-sink separation $t_s$. The lattice data is well described by
	the two-state fit curves. The gray band indicates the result of the two-state fit
	at the limit of $t_s\to\infty$.}
	\label{fig:excited_state}
\end{figure}

{\bf Treatment of the excited-state effects} --
We calculate the soft function at four different source-sink separations with $t_s/a=6,8,10,12$.
In Fig.~\ref{fig:excited_state}, we show the case with $\{P^z,b_\perp,m_\pi\}=\{3\frac{2\pi}{L},4a,\mbox{827 MeV}\}$ as an example.
The lattice results for various $t_s$ are shown together with the two-state fit curves and the ground state contribution (gray band). The same fit range works well for other values of $\{P^z,b_\perp,m_\pi\}$.

\begin{figure}[htb]
	\includegraphics[width=0.48\textwidth,angle=0]{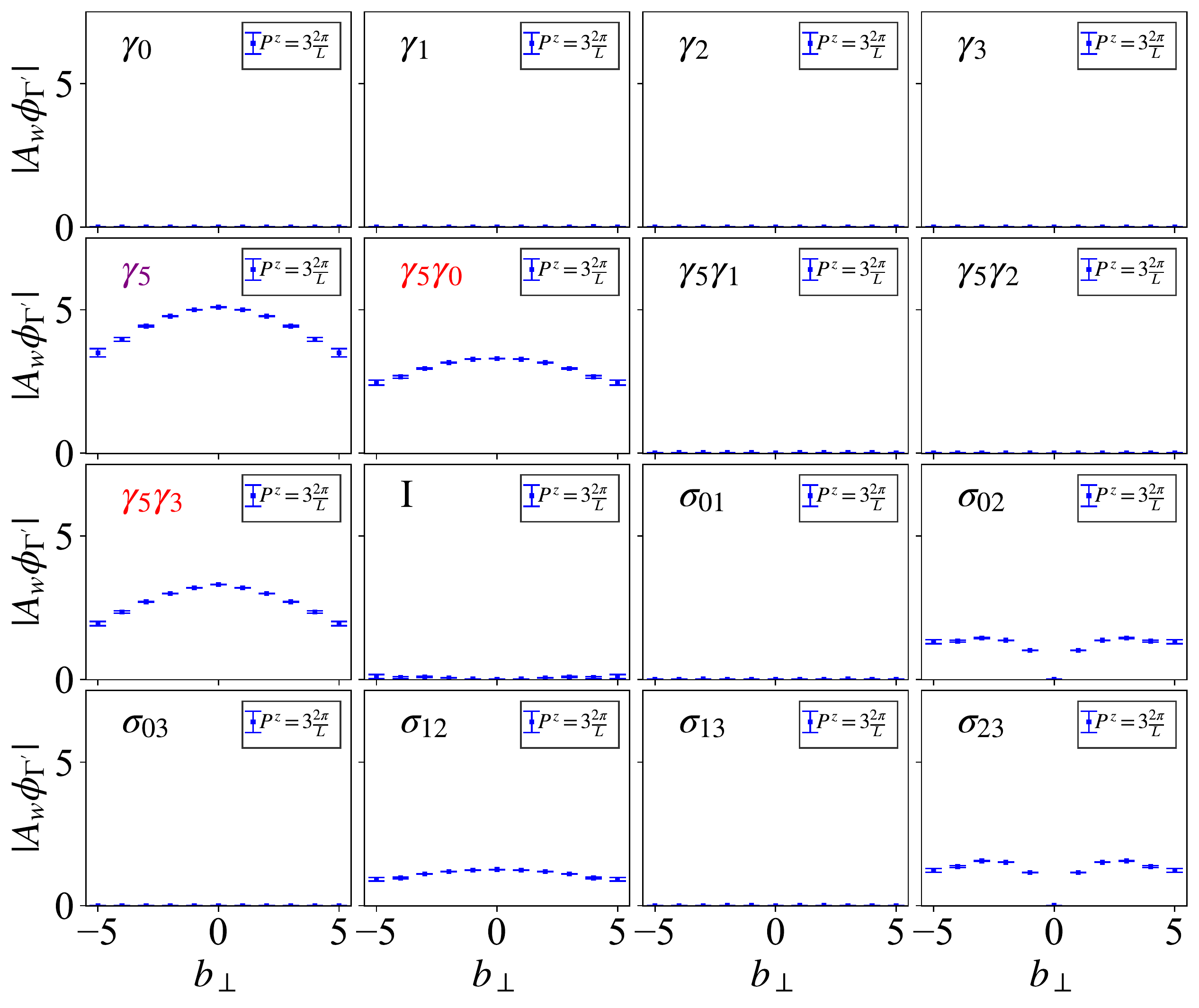}
	\includegraphics[width=0.48\textwidth,angle=0]{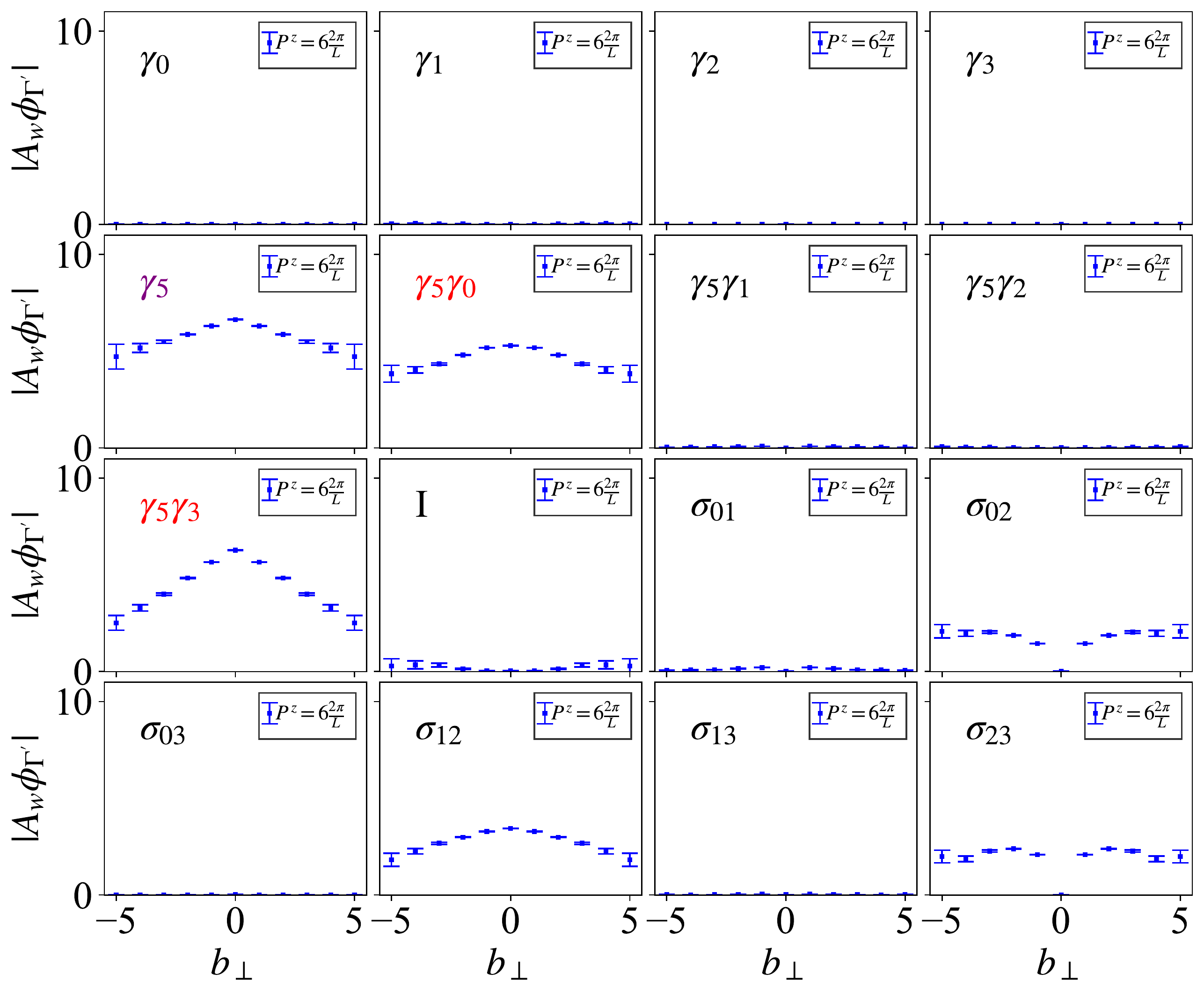}
	\caption{The product $|A_w(P^z)\,\phi_{\Gamma'}(b_\perp,l,P^z)|$ as a function of $b_\perp$ for various $\Gamma'$. Here, we show the results calculated at smallest momentum $P^z=3\frac{2\pi}{L}$ (top panel) and the largest one $P^z=6\frac{2\pi}{L}$ (bottom panel), both at $m_\pi=827$ MeV.
The wave functions $|\phi_{\gamma_5\gamma_0}|$
and $|\phi_{\gamma_5\gamma_3}|$ contain the LT contribution and are highlighted with red color. The largest HT contribution
$|\phi_{\gamma_5}|$ (purple color) significantly overwhelms the LT ones at $P^z=3\frac{2\pi}{L}$.
As $P^z$ increases, its relative size with respect to the LT ones decreases, which is consistent with the expectation.}
	\label{fig:TMDWF_matrix}
\end{figure}

{\bf Removal of the HT contamination} --
In Fig.~\ref{fig:TMDWF_matrix}, we show for various $\Gamma'$ 
the product of the renormalized wave function $|\phi_{\Gamma'}(b_\perp,l,P^z)|$ 
and the overlap amplitude $|A_w(P^z)|$ defined in Eq.~(\ref{eq:C_3pt}).
The wave function is renormalized using the ratio scheme as described in the paper.
The benefit to keep $|A_w|$
is to reduce the statistical uncertainties and thus to favor a better comparison. Note that given each momentum $P^z$, $|A_w|$ is
a universal factor for various $\Gamma'$ and thus does not affect the comparison.
Only the wave functions $|\phi_{\gamma_5\gamma_0}|$
and $|\phi_{\gamma_5\gamma_3}|$ contain the LT contribution, while all the others also contain a HT contribution.
We obtain from the figure that some HT contributions have comparable size to the LT ones.
We, thus, identify the four largest HT contributions with $\Gamma'=\gamma_5,\sigma_{02},\sigma_{12},\sigma_{23}$.
The next step to remove the large HT effects is to form appropriate combinations of $F_\Gamma$ with $\Gamma=I,\gamma_5,\gamma_\perp,\gamma_5\gamma_\perp$.

\begin{figure}[htb]
        \includegraphics[width=0.48\textwidth,angle=0]{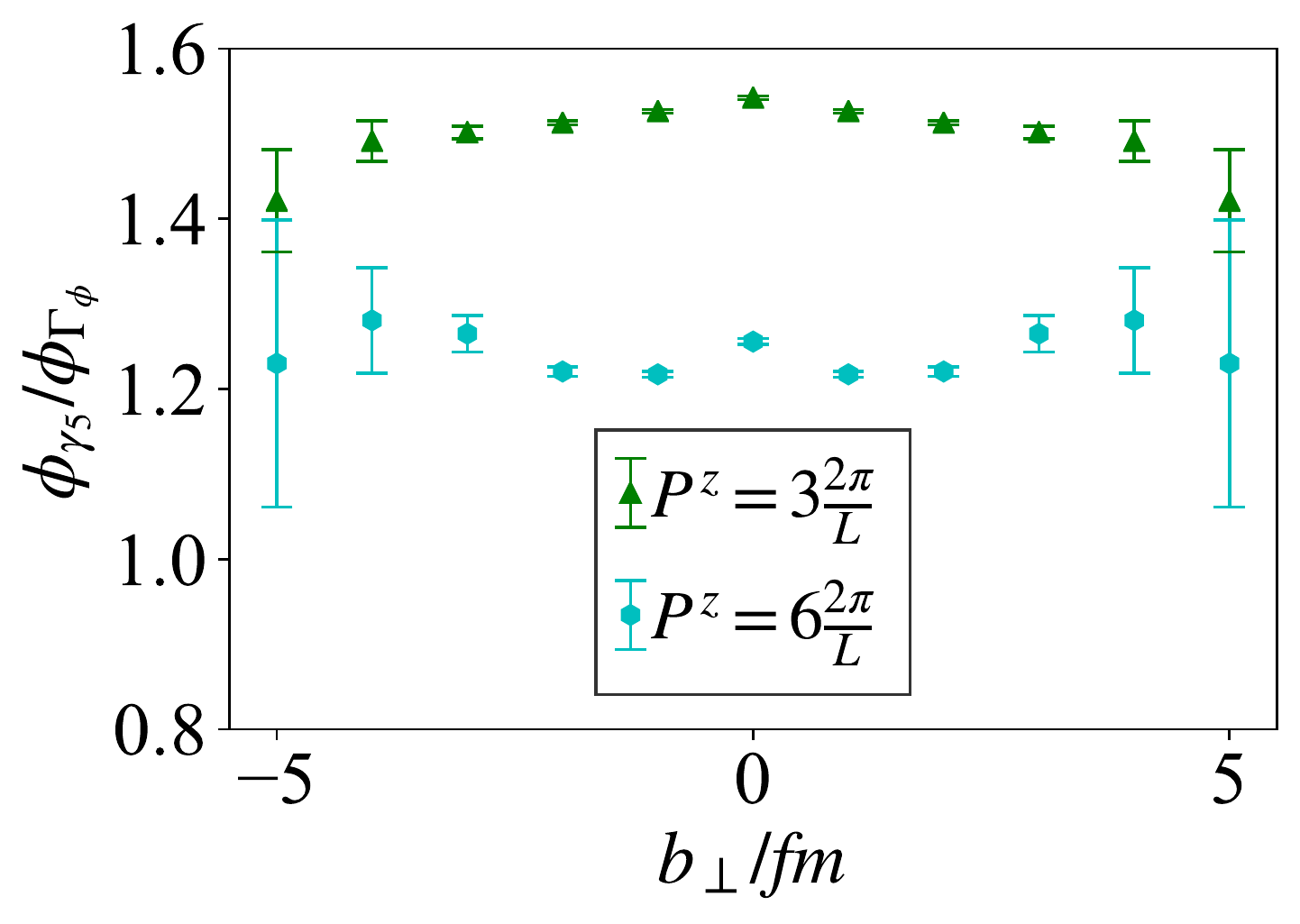}
        \caption{A ratio between the largest HT contribution with $\Gamma'=\gamma_5$ and the LT contribution 
with $\Gamma'=\Gamma_\phi=\gamma_5\gamma_0$ at $m_\pi=827$ MeV. The comparison suggests that when $P^z$ becomes sufficiently large, the LT contribution may finally become dominant.}  
        \label{fig:ratio_HT_LT}
\end{figure}

In Fig.~\ref{fig:ratio_HT_LT}, we show a ratio between the largest HT contribution with $\Gamma'=\gamma_5$ and the LT contribution
with $\Gamma'=\Gamma_\phi=\gamma_5\gamma_0$. (We do not use the LT contribution with $\gamma_5\gamma_3$ because at large $P^z$ some discrepancies between 
$\gamma_5\gamma_0$ and $\gamma_5\gamma_3$ are found. It has been pointed out earlier in the paper that $\Gamma_\phi=\gamma_5\gamma_0$ is a better choice to avoid the operator mixing.) Fig.~\ref{fig:ratio_HT_LT} exhibits a tendency that 
as $P^z$ increases the ratio decreases, which is consistent with the prediction from LaMET.

\begin{table}[htb]
	\begin{center}
		\begin{tabular}{|c|rrrrrr|}
			\hline
			\diagbox{$\Gamma'$}{$\Gamma$}& \multicolumn{1}{c}{$I$} & \multicolumn{1}{c}{$\gamma_1$} & \multicolumn{1}{c}{$\gamma_2$}
				&\multicolumn{1}{c}{$\gamma_5$} & \multicolumn{1}{c}{$\gamma_5 \gamma_1$} & \multicolumn{1}{c|}{$\gamma_5\gamma_2$} \\
			\hline
			$\gamma_5\gamma_0$	&1/4	   &1/4		  &1/4	      &$-1/4$		&1/4			 &1/4		\\
			$\gamma_5\gamma_3$	&1/4	   &1/4		  &1/4	      &$-1/4$		&1/4			 &1/4		\\
			\hline
			$\gamma_5$		&$-1/4$	   &1/4		  &1/4	      &$-1/4$		&$-1/4$			 &$-1/4$		\\
			$\sigma_{02}$		&$-1/4$	   &$-1/4$	  &1/4	      &$-1/4$		&1/4			 &$-1/4$		\\
			$\sigma_{12}$		&$-1/4$	   &1/4		  &1/4	      &$-1/4$		&$-1/4$			 &$-1/4$		\\
			$\sigma_{23}$		&$-1/4$	   &$-1/4$	  &1/4	      &$-1/4$		&1/4			 &$-1/4$		\\
			\hline
		\end{tabular}	
	\end{center}
	\caption{ Leading-order hard kernel $H_{\Gamma\Gamma'}^0$ using the Euclidean gamma matrices as input. All values from the table should be multiplied by a factor of $1/N_c$.}	
	\label{tab:LO_kernel}
\end{table}

\begin{figure}[htb]
	\includegraphics[width=0.48\textwidth,angle=0]{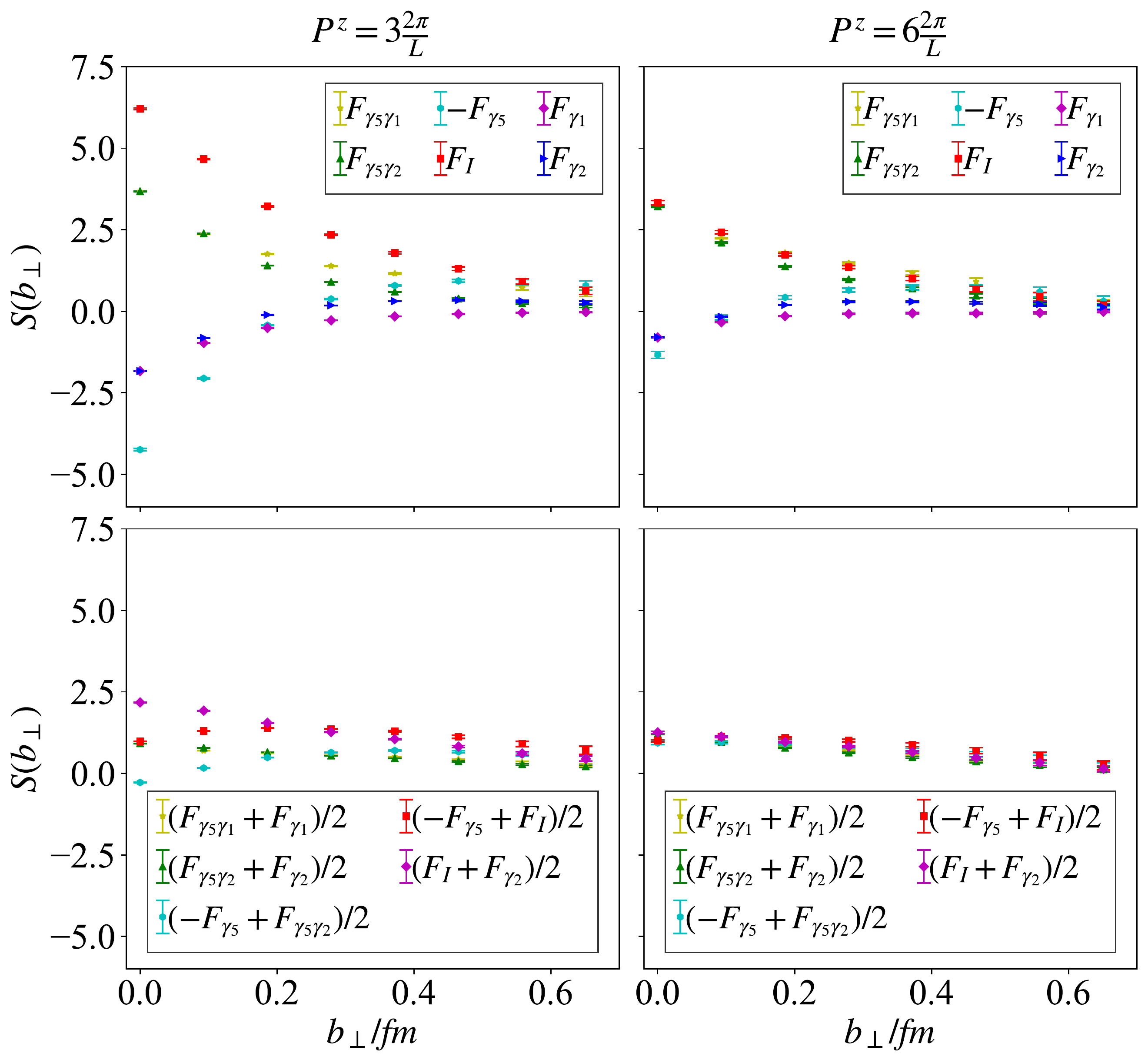}
	\caption{Examination of the convergence of the soft function when $P^z$ increases. The results are shown at $m_\pi=827$ MeV.}
	\label{fig:convergence}
\end{figure}

In Table~\ref{tab:LO_kernel}, the values of the LO hard kernel $H_{\Gamma\Gamma'}^0=\frac{1}{16N_c}\operatorname{Tr}(\Gamma\Gamma'\Gamma\Gamma')$ are shown for $\Gamma=I,\gamma_5,\gamma_\perp,\gamma_5\gamma_\perp$ associated with $\Gamma'=\gamma_5\gamma_0,\gamma_5\gamma_3$ (LT) and
 $\Gamma'=\gamma_5,\sigma_{02},\sigma_{12},\sigma_{23}$ (HT).
Using the information from Table~\ref{tab:LO_kernel}, we construct the five improved pion MEs given in Eq.~(\ref{type}).

In Fig.~\ref{fig:convergence}, we examine the convergence of the soft function when $P^z$ increases. For a comparison, we show the results using the $F_\Gamma$
with $\Gamma=I,\gamma_5,\gamma_\perp,\gamma_5\gamma_\perp$ in the upper panel
and the results using the improved pion MEs in the lower panel.
From left to right, the momentum $P^z$ increases from $3\frac{2\pi}{L}$ to $6\frac{2\pi}{L}$ and better convergence
is observed at larger momentum.
Unfortunately, due to the large HT contamination, even at $P^z=6\frac{2\pi}{L}$ $F_\Gamma$ with various $\Gamma$ still show a strong variation.
On the other hand, the improved pion MEs show
much better convergence, demonstrating that the HT effects are reduced significantly.

\begin{figure}
	\includegraphics[width=0.48\textwidth,angle=0]{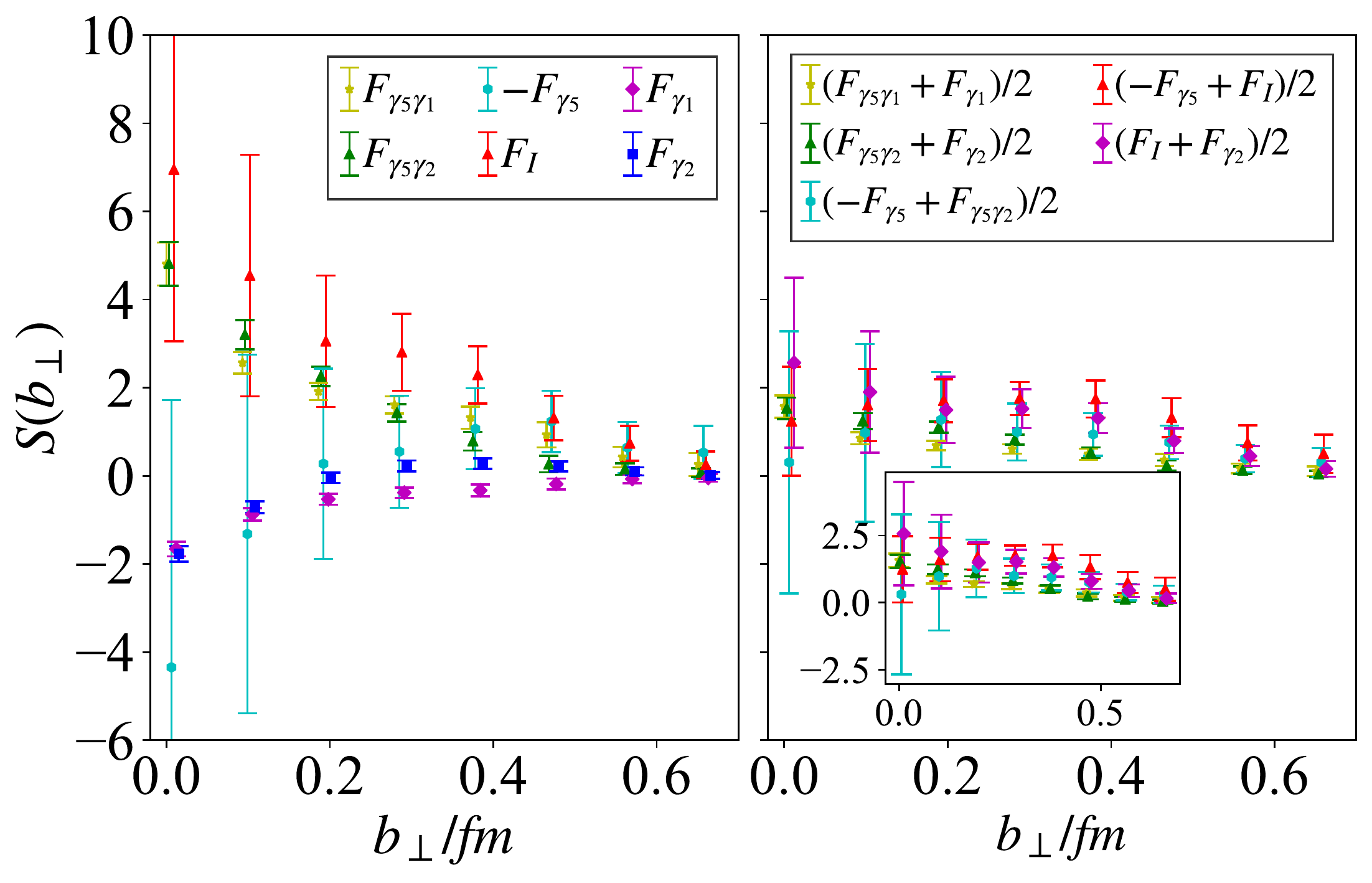}
	\caption{A figure similar to Fig.~\ref{fig:higher_twist_contamination} but at the lightest pion mass $m_\pi=350$ MeV.}
	\label{fig:higher_twist_contamination_lightest_pion}
\end{figure}

In Figs.~\ref{fig:higher_twist_contamination} and \ref{fig:convergence}, the results are calculated at $m_\pi=827$ MeV. To demonstrate that the improved MEs work well at
 different pion masses, in Fig.~\ref{fig:higher_twist_contamination_lightest_pion} we present a figure similar to Fig.~\ref{fig:higher_twist_contamination} but at the lightest pion mass $m_\pi=350$ MeV. Although the statistical uncertainties become much larger, the conclusions of the paper are not altered qualitatively.

\end{document}